\documentclass{kluwer}    

\newdisplay{guess}{Conjecture}

\def\spose#1{\hbox to 0pt{#1\hss}}
\def\lta{\mathrel{\spose{\lower 3pt\hbox{$\mathchar"218$}}
     \raise 2.0pt\hbox{$\mathchar"13C$}}}
\def\gta{\mathrel{\spose{\lower 3pt\hbox{$\mathchar"218$}}
     \raise 2.0pt\hbox{$\mathchar"13E$}}}
\def\etal{{\it et al.\ }}

\begin{document}                                                                                   
\begin{article}
\begin{opening}         
\title{The Physical Relation between Age and Metallicity in Galaxies} 
\author{Uta \surname{Fritze -- v. Alvensleben}}  
\runningauthor{Uta Fritze -- v. Alvensleben}
\runningtitle{The Physical Relation between Age and Metallicity in Galaxies}
\institute{Universit\"atssternwarte G\"ottingen}

\begin{abstract}
We present unified chemical and spectrophotometric evolutionary synthesis models that allow to describe composite stellar populations in a chemically consistent 
way. 
Keeping track of the ISM abundance at birth of each star and using sets of input physics (stellar yields, tracks, spectra) for different metallicities we account for the increasing initial metallicity of successive stellar generations. 
For any galaxy type its Star Formation History determines the physical relation between the ages and metallicities of its stars. As compared to models using solar metallicity input physics only, differences are significant, both for the ISM enrichment evolution and for the spectrophotometric evolution, and they increase with lookback time. Results are used to interprete Damped Ly$\alpha$ Absorbers. Knowledge of the chemical evolution of the ISM allows to predict abundances of stars, star clusters, and tidal dwarf galaxies that may form when spirals merge at any redshift.  

\end{abstract}

\end{opening}           
\vspace{-.4cm}
\section{Introduction}  
\vspace{-.3cm}
There are basically two kinds of stellar populations, simple and composite ones. {\large\bf S}imple {\large\bf S}tellar {\large\bf P}opulations ({\bf SSP}s) are single age single metallicity populations like star clusters. For normal galaxies, there is broad agreement among all galaxy evolution groups that the different spectral types of galaxies (E, S0, Sa, ..., Sd) are described by {\large\bf S}tar {\large\bf F}ormation {\large\bf H}istories ({\bf SFH}s) with characteristic timescales for the transformation of gas into stars that range from $\sim 1$ Gyr for E through a Hubble time for Sd galaxies. Due to their SFHs extending over more than the lifetime of the most massive stars, all types of galaxies, however, are {\large\bf C}omposite {\large\bf S}tellar {\large\bf P}opulations ({\bf CSP}s) with their stars comprising finite ranges both in age and metallicity. 

\noindent
In evolutionary models for SSPs, the well-known {\bf age -- metallicity degeneracy} describes the fact that an SSP of given age and metallicity is indistinguishable in optical broad band colors and stellar continuum shape from another one with half its age and 3 times its metallicity (cf. Worthey's (1994) 3/2 - rule). Stellar absorption indices and less age dependent NIR colors help to solve this ambiguity and disentangle the effects of age and metallicity. 
The CSP of any type of galaxy can be decribed by a superposition of spectra or luminosity contributions to any wavelength or filter band of SSPs of various ages and metallicities. I.e., any CSP can be expanded into a series of SSPs. 

\vspace{-.2cm}
\section{The Relation between Age and Metallicity in Normal Galaxies}
\vspace{-.2cm}
For normal galaxy types, however, not any arbitrary combination of ages and metallicities is realised. Instead, 
{\bf 1. there is a physical relation between age and metallicity in normal galaxy types} (cf. the age -- metallicity -- relation reported for the Milky Way by Twarog (1980) and others since then). {\bf 2. This relation is determined by the SFH of the respective galaxy type.} 
3. There is an intrinsic scatter around this relation. E.g. G-dwarf stars in the solar neighborhood show a metallicity distribution ${\rm -0.9 \leq [Fe/H] \leq +0.3}$ (Rocha -- Pinto \& Maciel 1998). 
Bulge stars in Baade's window feature a range of more than a factor 10 in metallicty ${\rm -1.3 \leq [Fe/H] \leq +0.5}$ (Mc William \& Rich 1994, Sadler \etal 1996), and 
the same is true for stars in elliptical galaxies.

The evidence for extended metallicity distributions in galaxies as well as the increasing importance of subsolar metallicities both towards later galaxy types in the local Universe and for less evolved galaxies at high redshifts clearly shows the need for a chemically consistent treatment of evolving galaxy populations (cf. F.-v.A. 1999). {\bf {\large C}hemically {\large C}onsistent (= cc) evolutionary synthesis models} combine the description of the chemical evolution of the ISM -- by solving a modified form of Tinsley's (1972) equations -- and of the spectrophotometric evolution of the stellar population -- by following all the stars that are formed on their respective paths through the HR diagram and attributing appropriate spectra to them. Keeping track of the ISM abundance at birth of each star and using sets of input physics (stellar yields, tracks, spectra) for different metallicities ${\rm 10^{-4} \leq Z \leq 2 \cdot Z_{\odot}}$ we account for the increasing initial metallicity of successive stellar generations. 
\vspace{-0.2cm}

\subsection{Metallicity Observations in Galaxies}
\vspace{-0.2cm}
Galaxy metallicities come in two flavors,  ISM abundances and stellar metallicities. For galaxies in the local Universe, an upper limit to the average ISM abundance is derived from HII region spectral emission lines. High resolution spectra of QSO absorption lines, in particular those of Damped Ly$\alpha$ systems (= {\bf DLA}s), reveal precise neutral gas phase abundances up to redshifts ${\rm z \gta 4}$, i.e. over more than 90 \% of the age of the Universe (e.g. Pettini \etal 1999). DLAs at high redshift are the progenitors of present-day massive galaxies, most probably of normal spirals. Average stellar metallicities, on the other hand, are derived from absorption lines like Mg$_2$, Fe52/53, etc. for local galaxies and from UV OB-stellar wind lines redshifted into the optical for high redshift galaxies (cf. Trager \etal 1997).   
\vspace{-0.2cm}

\subsection{Metallicities in Galaxy Models}
\vspace{-0.2cm}
Chemically consistent evolutionary synthesis models show how the stellar metallicity distributions in different galaxy types build up with time and, after a Hubble time, produce stellar metallicity distributions similar to those observed. They also show that the stronger SF varies with time the stronger can be the difference between stellar and ISM abundances, and that the luminosity-weighted stellar metallicities depend on the wavelength of observation (M\"oller \etal 1997a).
\vspace{-0.2cm}

\subsection{Closed Box Models}
\vspace{-0.2cm}
While, clearly, all hierarchical formation scenarios predict galaxies to build up gradually, they do not 
agree on the 
redshift evolution of the average mass (in gas and stars) of a typical present-day galaxy. 
The closed box models we use for normal galaxies to keep the number of free parameters low may to some degree be justified by observations. E.g., the luminosities of galaxies at ${\rm z \sim 1 - 2}$, i.e. at less than half their present ages, indicate masses roughly comparable to (and not $\sim \frac{1}{10}$ of) those today. Rotation curves for spirals out to ${\rm z \sim 1}$ (Vogt 2000) suggest the same. The kinematics of DLA absorbers at ${\rm z \sim 2 - 3}$ are consistent with rotation at ${\rm v_{rot} \sim 200~km~s^{-1}}$ and indicate that (proto-)spirals at ${\rm z \sim 2 - 3}$ already have masses comparable to those of local spirals (Prochaska \& Wolfe 1998). With the assumption of closed box models, cc evolutionary synthesis models yield the time evolution of both ISM abundances and stellar abundances as a function of SFH and, hence, show the physical relation between age and both types of metallicities in normal galaxy types. With a cosmological model as specified by the parameters (${\rm H_o,~\Omega_o,~\Lambda_o}$) and a redshift of galaxy formation ${\rm z_f}$, our cc evolutionary synthesis models give the redshift evolution of luminosities, colors, spectra (M\"oller \etal 1997b, 2000 {\sl in prep.}), and ISM abundances.
\vspace{-0.4cm}

\section{Enrichment History of Spiral Galaxies}
\vspace{-0.2cm}
The ISM enrichment history of our cc model galaxies can be studied in terms of individual abundances for a series of elements with various nucleosynthetic origins, as PNe, SNII, SNIa/b. It is compared to the observed redshift evolution of DLA abundances in Lindner \etal 1999. We show that over $\gta 90$ \% of the age of the Universe, DLA abundances are well described by our normal spiral models from Sa all through Sd 
with the earlier spiral types Sa, Sb dropping out of DLA samples towards lower redshifts ${\rm z \lta 1.5}$ because of their low gas content\footnote{Spectroscopic implications of this suggested change in the DLA galaxy population are well consistent with and, in fact explain, the non-detection of DLA galaxies in deep searches (cf. F.-v.A. \etal 1999).}. Without any free parameters or scaling, models with SFHs that bring the spectral evolution in agreement with spiral galaxy observations over the redshift range ${\rm 0 \leq z \leq 1}$ directly produce absolute ISM abundances that bridge the gap from DLA absorbers at ${\rm z = 0.4 ~\dots ~4.4}$ to HII region observations of local spiral types Sa ... Sd. The weak overall increase in DLA abundances from ${\rm z > 3}$ to ${\rm z \lta 1}$ is a consequence of the long SF timescales in spirals. The abundance scatter at fixed redshift reflects the range of SF timescales from 2 Gyr in Sa through 15 Gyr in Sd galaxies. 
While primordial infall at a constant rate increasing the galaxy mass by up to 50 \% since ${\rm z = 5}$ would not affect our results, a much stronger or time-dependant rate would. 

Knowledge of the ISM abundance evolution allows to predict the metallicity of stars, star clusters, and tidal dwarf galaxies that may form during spiral -- spiral mergers and their accompanying strong starbursts at various times or redshifts. Our metallicity predictions for star clusters and tidal dwarf galaxies forming in recent spiral -- spiral mergers (F.-v.A. \& Gerhard 1994a,b, F.-v.A. \& Burkert 1995) were well confirmed by spectroscopy of several of these objects (Schweizer \& Seitzer 1998, Whitmore \etal 1999, Duc \& Mirabel 1999), giving independent justification for the relation between age and ISM metallicity obtained from our models.

\end{article}

\vspace{-.4cm}
\begin{thebibliography}{}

\bibitem[]{} Duc, P.-A., Mirabel, I. F., 1999, IAU Symp. 186, 61
\bibitem[]{} Fritze -- v. Alvensleben, U., 1999, in {\sl Spectrophotometric Dating of Stars and Galaxies}, eds. 		I. Hubeny \etal, ASP Conf. Ser. 192, 273
\bibitem[]{} Fritze -- v. Alvensleben, U., Burkert, A., 1995, A\&A 300, 58
\bibitem[]{} Fritze -- v. Alvensleben, U., Gerhard, O. E., 1994a, b, A\&A 285, 751 + 775
\bibitem[]{} Fritze - v. Alvensleben, U., Lindner, U., M\"oller, C. S., 1999, in {\sl Chemical
	Evolution from Zero to High Redshift}, ed. J. Walsh, Springer, p. 256
\bibitem[]{} Lindner, U., Fritze - v. Alvensleben, U., Fricke, K. J., 1999, A\&A 341, 709
\bibitem[]{} McWilliam, A., Rich, R. M., 1994, ApJS 91, 749 
\bibitem[]{} M\"oller, C. S., Fritze - v. Alvensleben, U., Fricke, K. J., 1997a, A\&A 317, 676
\bibitem[]{} M\"oller, C. S., Fritze - v. Alvensleben, U., Fricke, K. J., 1997b, in {\sl HDF Symposium}, eds. M. Livio, S. M. Fall, P. Madau, STScI, p. 53
\bibitem[]{} Pettini, M., Ellison, S. L., Steidel, C. C., Bowen, D. V., 1999, ApJ 510, 576
\bibitem[]{} Prochaska, J. X., Wolfe, A. M., 1998, ApJ 507, 113
\bibitem[]{} Rocha - Pinto, H. J., Maciel, W. J., 1998, A\&A 339, 791
\bibitem[]{} Sadler, E. M., Rich, R. M., Terndrup, D. M., 1996, AJ 112, 171
\bibitem[]{} Schweizer, F., Seitzer, P., 1998, AJ 116, 2206
\bibitem[]{} Tinsley, B. M., 1972, {\sl Fun. Cosmic Physics} 5, 287
\bibitem[]{} Trager, S. C., Faber, S. M., Dressler, A., Oemler, A., 1997, ApJ 485, 92
\bibitem[]{} Twarog, B. A., 1980, ApJ 242, 242
\bibitem[]{} Vogt, N. P., 2000, ASP Conf. Ser. 197, 435
\bibitem[]{} Whitmore, B. C., Zhang, Q., Leitherer, C., \etal 1999, AJ 118. 1551
\bibitem[]{} Worthey, G., 1994, ApJS 95, 107

\end{thebibliography}
\end{document}